\documentclass{PoS}
\title{Is covariant star product unique?}

\ShortTitle{Covariant star product}

\author{\speaker{Dmitri Vassilevich}%
\\
       CMCC - UF ABC (Brazil)\\
       E-mail: \email{dvassil@gmail.com}}

\abstract{We give a nontechnical introduction to the problem of non-uniqueness
 of star products and describe a covariant resolution of this problem.
Some implications (e.g., for noncommutative gravity) and further prospects
are discussed.}

\FullConference{Corfu Summer Institute on Elementary Particles and Physics - Workshop on Non Commutative Field Theory and Gravity\\
                September 8 - 12, 2010\\
                Corfu, Greece}

\begin{document}

\section{Introduction}
The arena for a noncommutative field theory is a noncommutative manifold, which is
defined by replacing the usual point-wise product of functions 
by a noncommutative star product. This star product depends on a deformation parameter
$h$, and the first term in the $h$-expansion is essentially the Poisson bracket
of functions. It is a known result by Kontsevich \cite{Kontsevich} that the Poisson
structure defines the star product up to an equivalence. This, however, does not
imply that the star product for a given Poisson structure is unique. On the opposite,
the family of equivalence transformations is very rich, it depends on an infinite
number of arbitrary functions. Consequently, the ambiguity in definition of the star
product has an infinite functional dimension. This ambiguity is physically relevant
as equivalent star products lead, in general, to inequivalent field theories.
Therefore, before going on with constructing a field theory on noncommutative
space one has to decide which of the star products has to be used in this construction.

It is natural to request that the algebra equipped with the star product has a
number of derivations (represented by covariant derivatives)
which satisfy the usual Leibniz rule. However, since derivative
of a scalar is a vector, and a product of two vectors is a rank-two tensor,
to impose the Leibniz rule one has to extend the star product to all tensor fields.
Tensors are characterized by their transformation properties with respect to 
diffeomorphisms, these properties should be respected by the deformation. Consequently,
the star product has to be covariant. This is why we address here the problems of
covariance and uniqueness simultaneously.

The paper is organized as follows. In Sec.\ \ref{sec-non} we introduce
star products and describe ambiguities in their definition. Covariant
star products are treated in a very short Sec.\ \ref{sec-cov}. Our main
result \cite{Vassilevich:2010he}  is a set of natural conditions which allow to remove 
the ambiguity almost completely (Sec.\ \ref{sec-red}). Some ambiguity,
however, remains. In Sec.\ \ref{sec-cla} we show that it leads to classically
equivalent noncommutative field theories. In Sec.\ \ref{sec-wha} we discuss
possible extension of the results (to Yang-Mills symmetries, Poisson
manifolds, strict deformation quantization) as well as implications for
noncommutative gravity.
\section{Non-uniqueness of star products}\label{sec-non}
In the approach adopted here, which goes back to the seminal papers \cite{BFFLS1,BFFLS2} (see also \cite{DS}),
the star product is defined through a noncommutative deformation of the algebra $\mathcal{A}$.
For simplicity, in this section we suppose that $\mathcal{A}$ is the algebra 
$C^\infty(\mathbb{R}^n)$ of smooth functions on $\mathbb{R}^n$. Next, we replace 
$\mathcal{A}$ by an algebra of \emph{formal power series} $\mathcal{A}[[h]]$ of the
deformation parameter $h$, i.e., of elements of the form
\begin{equation}
a=a_0+ha_1+h^2a_2+\dots \label{aform}
\end{equation}
Here, each $a_j$ is a smooth function. A new product is then defined as
\begin{equation}
a \star b = a \cdot b +\sum_{r=1}^\infty h^r C_r(a,b) \,,\label{asb}
\end{equation}
where each $C_r$ is a bidifferential operator. This means, for one argument fixed
$C_r$ is a differential operator acting on the other argument. We stress, that 
the series (\ref{aform}) and (\ref{asb}) are formal, i.e., no convergence is assumed.

In physical applications the deformation parameter $h$ is usually taken 
to be imaginary.

It is required that the star product is associative, $(a\star b)\star c=a\star (b\star c)$.
In the first order of $h$ this yields
\begin{equation}
C_1(a,b)-C_1(b,a)=2\{ a,b\} \,,\label{stabr}
\end{equation}
where $\{ a,b\}$ is a Poisson bracket,
\begin{equation}
\{ a,b\} =\omega^{\mu\nu}\partial_\mu a \partial_\nu b \label{PbPs}
\end{equation}
with some Poisson bivector $\omega^{\mu\nu}$. 

Another common requirement is the Moyal symmetry,
\begin{equation}
C_r(a,b)=(-1)^r C_r(b,a) \,.\label{Moysym}
\end{equation}
Then $C_1(a,b)$ is defined to be a Poisson bracket. One says that the star
product (\ref{asb}) is a deformation of the point-wise product in the direction
of $\omega^{\mu\nu}$. The problem is to find all star products for a given Poisson
structure. Up to second order, a solution may be found by straightforwardly solving
the associativity condition,
\begin{eqnarray}
&&a\star b= ab + h\omega^{\mu\nu}\partial_\nu a \partial_\nu b +
h^2 \left( \frac 12 \omega^{\mu\nu}\omega^{\rho\sigma}(\partial_\mu\partial_\rho a)
(\partial_\nu\partial_\sigma b)\right. \nonumber\\
&&\qquad\qquad\qquad \left.+ \frac 13 \omega^{\mu\nu}\partial_\nu \omega^{\rho\sigma}
((\partial_\mu \partial_\rho a)(\partial_\sigma b) - (\partial_\rho a)(\partial_\mu
 \partial_\sigma b))\right)+O(h^3)\,.
\label{sto2}
\end{eqnarray}

The existence of a solution to the deformation problem to all orders for an arbitrary Poisson
structure was demonstrated by Kontsevich \cite{Kontsevich}, and the solution is not
unique. If $\star$ is a product corresponding to some fixed Poisson bivector
$\omega^{\mu\nu}$, then one can define another  product $\star'$ by
\begin{equation}
a\star'b = D^{-1} (D a \star D b),\label{gatr}
\end{equation}
where
\begin{equation}
D=1+hL \,.\label{Dtau}
\end{equation}
and $L$ being a formal differential operator (a
sum of differential operators of arbitrary order with arbitrary polynomial dependence
on $h$). Clearly, $\star'$ is also an associative product corresponding to the same
Poisson bivector\footnote{Actually, Kontsevich \cite{Kontsevich} demonstrated a more
refined statement. One defines a \emph{formal Poisson structure}, which is, roughly
speaking, a Poisson structure which is a formal polynomial in $h$. The classes of equivalence
of the star products are in one-to-one correspondence with equivalence classes of 
formal Poisson structures.}. The transformation $\star\to\star'$ was called a gauge
transformation in \cite{Kontsevich}.

The products $\star$ and $\star'$ are equivalent and correspond to isomorphic algebras,
but they are \emph{different}. As we shall see below, field theory models constructed
with equivalent star products are in general non-equivalent. Therefore, one has an ambiguity
in the choice of the star product corresponding to the ambiguity in the choice of $D$
in (\ref{gatr}). It is easy to understand that the freedom in $D$ has an infinite functional
dimension (infinite number of parameters per point in $\mathbb{R}^n$). 

The situation in noncommutative theories differs drastically from that in, say, general
relativity. The metric tensor defines the whole (pseudo-)riemannian structure of the manifold,
while to define a star product\footnote{Actual calculation of a star product may also
be problematic. The formulae in \cite{Kupriyanov:2008dn} work till the 5th order
in $h$, and beyond this order little is known.}
 (which is the basic structure on a noncommutative space) one
has to specify an infinite number of fields contained in $D$ in addition to a Poisson tensor.

\section{Covariant star products}\label{sec-cov}
The product (\ref{sto2}) is not covariant at the second order of the expansion
in $h$ since it contains usual derivatives of the Poisson tensor. Over last years 
much progress has been achieved in covariantization of the Kontsevich procedure
\cite{Dolgushev,ACG} and in extending star products to differential forms
\cite{HoM,tenv,McCZ}, to Lie algebra valued differential forms \cite{Liev},
and in covariant holomorphic products \cite{Cornalba:1998kt}. 

In the symplectic case (non-degenerate  $\omega^{\mu\nu})$ situation is much better
understood. A covariant star product was suggested already in the seminal
paper \cite{BFFLS1}. An extension of this product to all tensor fields is
rather straightforward \cite{Vassilevich:2009cb,Vassilevich:2010he}.
The generic construction by Fedosov \cite{Fedosov1,Fedosov2}
is manifestly covariant. 

There is a general feature of (almost) all covariant approaches: to make the star
product covariant one needs a connection in addition to the Poisson or symplectic
structure.

\section{Reducing the gauge freedom}\label{sec-red}
From now on we restrict ourselves to symplectic manifolds.

To construct a star product we need some geometric data, which are going to be
a symplectic manifold $M$ with a flat symplectic connection $\nabla$. Let
$\omega_{\mu\nu}$ be a symplectic form (with the Poisson bivector $\omega^{\mu\nu}$
being its' inverse, $\omega_{\mu\nu}\omega^{\nu\rho}=\delta_\mu^\rho$). 
Let us choose a Christoffel symbol
on $M$ such that the symplectic form is covariantly constant,
\begin{equation}
\nabla_{\mu}\omega_{\nu\rho}=
\partial_\mu \omega_{\nu\rho}-\Gamma_{\mu\nu}^{\sigma}\omega_{\sigma\rho}
-\Gamma_{\mu\rho}^\sigma \omega_{\nu\sigma}=0.\label{cco}
\end{equation}
Therefore, ${M}$ becomes a Fedosov manifold \cite{GRS}. Let us suppose
that this connection is flat and torsion-free, i.e.,
\begin{equation}
 [\nabla_\mu,\nabla_\nu]=0.\label{flatcon}
\end{equation}
Locally, one can choose a coordinate system such that $\omega^{\mu\nu}=const.$ and $\Gamma_{\mu\nu}^\sigma=0$.
Such coordinates will be called the Darboux coordinates.

We are going to construct star products for arbitrary tensor fields
$\alpha_{n,m}\in TM^n \otimes T^*M^m\equiv T^{n,m}$. This means,
$\alpha_{n,m}$ has $n$ contravariant and $m$ covariant indices.
Next, we define a covariant Poisson bracket for the tensors
\begin{equation}
\{\alpha,\beta\}=\omega^{\mu\nu} \nabla_\mu\alpha \cdot \nabla_\nu\beta \label{pbr}
\end{equation}
possesing all the standard 
properties of a Possion bracket (antisymmetry, Jacobi identity, etc.). Besides,
\begin{equation}
\nabla \{ \alpha ,\beta \}=\{ \nabla\alpha,\beta\} +
\{\alpha,\nabla\beta\}.\label{Lbr}
\end{equation}

Now, we are ready to construct a star product, which has to be an associative deformation
of the point-wise product in the direction of the Poisson bracket (\ref{Lbr}), i.e.,
it is a product on $T[[h]]$ having the form (\ref{asb}) 
subject to (\ref{stabr}) (with $a,b$ replaced by $\alpha,\beta$). We are interested in
covariant products only, meaning that $T^{n_1,m_1}[[h]]\star T^{n_2,m_2}[[h]]\subset
T^{n_1+n_2,m_1+m_2}[[h]]$. Besides, we impose a few "natural" restrictions. Namely,
we require stability on covariantly constant tensors
\begin{equation}
\alpha\star\beta=\alpha \cdot \beta \quad
{\mbox{if}}\quad \nabla\alpha=0 \ \mbox{or}\ \nabla\beta=0\,,\label{stab}
\end{equation}
the Moyal symmetry
\begin{equation}
C_k(\alpha,\beta)=(-1)^k C_k(\beta,\alpha)\,\label{Moyas}\end{equation}
and that $\nabla$ is a derivation
\begin{equation}
\nabla \alpha\star\beta = (\nabla\alpha)\star\beta + \alpha\star
(\nabla\beta).\label{der}
\end{equation}

Let us discuss the meaning of (\ref{stab}) - (\ref{der}). The first condition (\ref{stab})
replaces the usual requirement that the unit element of algebra is not deformed. 
Physically, it means that slowly varying fields do not see noncommutativity. In applications,
the deformation parameter $h$ is imaginary. Therefore, the Moyal symmetry ensures
hermiticity of the star product. The last condition (\ref{der}), which nothing else than
the Leibniz rule, means that the star product transforms in a controlled way under
infinitesimal translations (those generators are the covariant derivatives). In a sense,
(\ref{der}) replaces locality of commutative products. 

The conditions (\ref{stab}) - (\ref{der}) appear to be very restrictive. 
One can show \cite{Vassilevich:2010he},
that if they are satisfied, the star product has the form
\begin{equation}
\alpha\star_N\beta = D^{-1}(D\alpha\star_RD\beta)\,,\label{starN}
\end{equation}
where
\begin{equation}
\alpha\star_R\beta =\sum_k \frac{h^k}{k!} \omega^{\mu_1\nu_1}_R
\dots \omega^{\mu_k\nu_k}_R (\nabla_{\mu_1} \dots \nabla_{\mu_k}
\alpha) \cdot (\nabla_{\nu_1}\dots \nabla_{\nu_k} \beta)\,,
\label{Rstar}
\end{equation}
which depends on a ``renormalized'' symplectic structure
\begin{equation}
\omega_R^{\mu\nu}=\omega^{\mu\nu}+h^2 \omega^{\mu\nu}_1+h^4\omega^{\mu\nu}_2+\dots
\label{Rom}
\end{equation}
with all correction terms $\omega^{\mu\nu}_j$ being covariantly constant,
\begin{equation}
 \nabla_\rho \omega^{\mu\nu}_j=0. \label{ccRom}
\end{equation}
The operator $L$ in the transformation $D$, see (\ref{Dtau}), must have the form
\begin{equation}
L = \sum_{k=2}^{\infty} L^{\mu_1 \dots \mu_k}\nabla_{\mu_1}\dots\nabla_{\mu_k}\,,
\label{tausum}
\end{equation}
and be a scalar (i.e., $L$ has to be proportional to a unit matrix in the tensor indices)
and all $ L^{\mu_1 \dots \mu_k}$ have to be covariantly constant,
\begin{equation}
\nabla_\nu L^{\mu_1\dots \mu_n}=0\,,\label{Lncc}
\end{equation}
and may contain odd powers of the deformation parameter only.

The product (\ref{Rstar}) is nothing else than a covariantization of the Moyal
product \cite{BFFLS1} extended to tensors \cite{Vassilevich:2009cb,Vassilevich:2010he}.

The proof \cite{Vassilevich:2010he} of this statement is rather hard and technical,
but we shall try to give here a rough idea how does it go. First, let us note that the
product (\ref{starN}) with an arbitrary form differential operator $D$ is a deformation
of the pointwise product of tensors in the direction of the Poisson brackets (\ref{pbr}).
Therefore, our task is to analyze restrictions on $D$. The "gauge freedom" contained
in $D$ is much larger than in the scalar case. Besides being a differential operator,
$L$ may be a complicated linear transformation of tensors. Mixing up tensors of
different degree is forbidden by the covariance. Therefore, $L^{\mu_1\dots \mu_n}=
\oplus_{n,m} L^{\mu_1\dots \mu_n}_{n,m}$, where $L^{\mu_1\dots \mu_n}_{n,m}$ is a restriction
to the tensors of degree $(n,m)$. Next, by using (\ref{stab}), one can show that
$L^{\mu_1\dots \mu_n}_{n,m}=I_{(n,m)}L^{\mu_1\dots \mu_n}_{0,0}$, where $I_{(n,m)}$
is the identity map. Further restrictions are then obtained by analyzing the Leibniz rule
(\ref{der}). The Moyal symmetry (\ref{Moyas}) is not very essential, though useful to
remove certain powers of the deformation parameter, as, e.g., odd powers of $h$ in
(\ref{Rom}). 

The ambiguity in the star products has been reduced enormously. Instead of an
infinity number of fields in the gauge operator $D$ we have just an infinite number
of constants. Nevertheless, some freedom still remains, and we like to discuss its'
meaning. The freedom encoded in (\ref{Rom}) is rather harmless. This is the price
to pay for working in formal setup. In physical application the series in (\ref{Rom})
must be summed up somehow. Only the "renormalized" value $\omega_R$ is of relevance,
while the way how we split it in the perturbation series is no more than a technical
device. The rest of the gauge freedom is discussed in the next section.

\section{Classical field theory}\label{sec-cla}
To understand consequences of the freedom encoded in remaining gauge transformations
one has consider the integration. There is a natural measure on symplectic manifolds
\cite{Felder:2000nc}
\begin{equation}
d\mu (x)=(\det (\omega^{\mu\nu}))^{-\frac 12}\, dx\,.
\label{intme}
\end{equation}
With respect to this measure, the star product $\star_R$ is closed, i.e.,
\begin{equation}
\int_{{M}} d\mu (x) \alpha_{\mu\nu\dots\rho}\star \beta^{\mu\nu\dots\rho}=
\int_{{M}} d\mu (x) \alpha_{\mu\nu\dots\rho}\cdot \beta^{\mu\nu\dots\rho}\,.
\label{clo}
\end{equation}
provided all indices are contracted in pairs (or, equivalently, if the integrand is
diffeomorphism invariant). Among the products $\star_N$ not all are closed.

Consider an action of a classical field theory on noncommutative space
equipped with the product $\star_N$
\begin{equation}
S=\int d\mu (x) P(f_i,\nabla)_{\star_N}\,,\label{SN}
\end{equation}
where $f_i$ are some fields, $P$ is a polynomial, where all products are $\star_N$ products.
We can rewrite $S$ as
\begin{equation}
S=\int d\mu (x)D^{-1}( P(Df_i,\nabla)_{\star_R})=
\int d\mu (x)P(Df_i,\nabla)_{\star_R}\,.\label{SSR}
\end{equation}
This means, that the replacement $\star_N$ by $\star_R$ is compensated by the transformation
$f_i\to Df_i$. Since the operator $D$ is invertible, the theories based on the two star products
are classically equivalent. Therefore, the remaining freedom is also harmless.
For physical applications, the star products related by gauge transformations with
covariantly constant coefficients in $L$ are equivalent.

The same calculations also shows why the theories based on gauge-equivalent star products
are, in general, non equivalent. If the tensors $L^{\mu_1\dots\mu_k}$ are not all covariantly
constant, one cannot delete the operator $D^{-1}$ is (\ref{SSR}).

\section{What is next?}\label{sec-wha}
\subsection{Yang-Mills gauge covariance}
For physical applications is important to have a star product which is covariant
with respect to the Yang-Mills gauge transformation. A product of this type was
constructed in \cite{Vassilevich:2007jg} by introducing a flat (gauge-trivial)
connection and covariantization of the Moyal product. That paper did not consider
the problem of uniqueness of gauge-covariant star products, and here we briefly comment
on how this can be achieved. First of all, introducing a matter field belonging to
certain representation of the gauge group requires an extension of the space of
function on which the star product is defined to a direct sum of all representations
appearing in tensor powers of the initial one. This is similar to introducing the
spaces $T^{n,m}$, see above. Then, one has to impose a set of "natural" restrictions
on the star product, with most important ones being the stability on covariantly
constant fields, the Leibniz rule, and the Moyal symmetry (exactly in parallel to
Eqs. (\ref{stab}) - (\ref{der})). Then one observes that the Moyal-type product
(\ref{Rstar}) with the gauge connection introduced above satisfies all the requirements.
Therefore, the problem is to find the restrictions of the Kontsevich transformations
$D$ which also respect that requirements. The calculations go like in the previous
case, but are much simpler as the gauge connection does not change the representation
of the gauge to which the field belongs (in contrast to the tensor degree). 
It is not surprising therefore that the final result looks precisely as before
with just the connection being suitably modified.

\subsection{Poisson manifolds}
On the physical grounds it is hard to explain why $\omega^{\mu\nu}$ has to be
non-degenerate. Therefore, one has to be able to work in a more general setting
of Poisson manifolds. This case is much more complicated than the symplectic one,
as, for example, no Moyal-type formula (see (\ref{Rstar}) exists. This is related
to non-existence of a connection such that $\nabla \omega^{\mu\nu}=0$ since covariant
constancy of $\omega$ implies that its' rank is a constant, which is not always
true on Poisson manifolds. This can be partially corrected by introducing the
so-called contravariant connection \cite{IV}.  
On the other hand, the construction of the universal star product \cite{ACG}
starts with an arbitrary connection, and one can be quite optimistic about
a possibility to extend this construction to tensors and to study the consequence
of the Leibniz rule and other restrictions which we imposed above in the
symplectic case. As we have already explained above, a honest implementation of
the Leibniz rule requires an extension to tensors.

\subsection{Beyond formality}
In physical applications it is desirable to sum up the series in deformation
parameter to obtain a star product between functions on a manifold rather than
between formal power series. One has to be able to give a numerical value to
$h$, and to make sense of the notions like "the characteristic scale of noncommutativity".
From mathematical point of view, summing up formal series is a step towards
using the full power of noncommutative geometry \cite{GVF}. The problem of
going beyond formality is extremely complicated. There are just a few manifolds
which are noncommutative in the strict sense. Therefore, it is probably more
reasonable to restrict ourselves to construcing new examples. The existence of a
flat torsionless symplectic connection (and of some more useful structures) is guaranteed
on rigid special K\"ahler manifolds \cite{rsK1,rsK2,rsK3} which seem to be ideal
candidates for (strict) noncommutative deformations\footnote{I am grateful to Paolo
Aschieri for drawing my attention to special K\"ahler geometries.}.

\subsection{Towards noncommutative gravity}
Proper realization of the diffeomorphism invariance is one of the most important
issues for constructing noncommutative gravity theories \cite{Szabo}. Of course,
having a diffeomorphism covariant star product helps a lot. In particular, one
can construct \cite{Vassilevich:2009cb}
noncommutative counterparts for all dilaton gravities in two dimension \cite{Grumiller:2002nm}
having a full diffeomorphism invariance group, which by itself looks as a rather
strong result. Since one of the models appeared to be integrable
to all orders of the noncommutativity parameter, analyzing the solutions allows to 
highlight some general problems arising in noncommutative gravity models. The dilaton
gravity models in the first order formulation contain a zweibein which can be represented
through a complex one-form $e_\mu$ so that the Lorentz transformations become 
multiplications with a local phase factor. In NC models this multiplications becomes
a star multiplication, which can be fixed to be a left star multiplication
\begin{equation}
\delta e_\mu =i\lambda \star e_\mu ,\qquad
\delta \bar e_\mu =-i\bar e_\mu \star \lambda,\label{Lotr}
\end{equation}
where $\lambda$ is a parameter. The transformation of complex conjugate zweibein
$\bar e_\mu$ involves right star multiplication. (This follows from the Moyal
symmetry and the fact that $h$ has to be considered imaginary). There is then a
unique metric
\begin{equation}
g_{\mu\nu}=\frac 12 (\bar e_\mu \star e_\nu +
\bar e_\nu \star e_\mu),\label{gmn}
\end{equation}
which is real, symmetric, invariant with respect to (\ref{Lotr}) and does not
contain derivatives. (The derivatives inside the star product do not count). Since
the star product is diffeomorphism covariant, the metric (\ref{gmn}) defines a
diffeomorphism invariant line element. Dilator gravity models also contain a dilaton,
a connection one-form, and two auxiliary fields which generate torsion constraints.
None of them will be important below. It is not hard to construct a noncommutative
2D dilaton gravity that is integrable and admits gauge-trivial solutions only
\cite{Vassilevich:2009cb}. In other words, up to a Lorentz transformation,
\begin{equation}
e_\mu =\partial_\mu E \,,\label{eE}
\end{equation}
where $E$ is a complex scalar function. By a suitable choice of the coordinate,
taking $E=x^1+ix^2$, the zweibein can be made the unit matrix. On the other hand,
there is a coordinate system where the symplectic structure is constant and the
connection is trivial. In such a system the star product becomes just the usual
Moyal product. The Moyal product of two unit zweibeins is a unit metric, and one
may conclude that the model describes a flat space geometry, as expected. The
last statement is however wrong as the zweibein and the symplectic structure
trivialize in \emph{two different} coordinate systems, in general. To understand
the consequences, let us consider an example. As we have just mentioned above,
one can choose a coordinate system to simplify the symplectic structure and
to make the star product the usual Moyal product,
\begin{equation}
\alpha \star \beta = \exp (i\theta (\partial_1^x \partial_2^y - \partial_2^x\partial_1^y)
\alpha (x)\beta(y)\vert_{y=x}. \label{Mstar}
\end{equation}
Let us take the function $E$ in most simple but yet nontrivial form, 
$E=\sin (x^1) + i \sin (x^2)$. Calculating the metric and corresponding 
invariants is an easy exercise, (one can also look up in\cite{Vassilevich:2010he}), 
showing that
in this example we are dealing with a very non-trivial geometry. The metric
even changes signature at some values of $x^1$ and $x^2$. The lesson one can
learn from this example is not that noncommutative gravity predicts signature
changes, but rather that the metric can behave in a very wild way. To avoid
this, one needs to impose a relation between the metric and the symplectic
structure. Interesting relations of that kind follow from the matrix models \cite{Steinacker:2010rh}.

Same problem exists in most of the approaches to noncommutative gravity,
specifically in that based on a fixed star product. Many such models are presented
in the review paper by Szabo \cite{Szabo}. For each of these models one should
either demonstrate that noncommutative corrections to a classical geometry
(say, Schwarzschild black hole) are physically equivalent independently
in which particular coordinate system (e.g., Schwarzschild or Eddington-Finkelstein)
the star product is Moyal, or to propose a principle which relates the metric
to the star product.

\section*{Acknowledgement}
I am grateful to Dorothea Bahns, Harald Grosse and George Zoupanos for organizing
such a wonderful workshop, for their support and stimulating atmosphere.
This work was supported in part by FAPESP and CNPq.

\end{document}